\documentclass[preprint,12pt]{elsarticle}

\usepackage{graphicx}
\usepackage{amssymb}
\usepackage{float}
\usepackage{lineno,hyperref}
\usepackage{amsmath,amsfonts,amsthm,amssymb,graphicx,subfigure,mathtools,tikz,hyperref,bm,blindtext}
\usepackage{diagbox}
\usepackage{geometry}
\usepackage[acronym]{glossaries}
\usepackage{multirow}
\usepackage{algorithm}
\usepackage{algpseudocode}
\usepackage{subfigure}

\usepackage{lineno}

\journal{Journal Name}

\begin{document}

\begin{frontmatter}


\title{Physics-informed neural networks for solving forward and inverse flow problems via the Boltzmann-BGK formulation}



\author[usst1,usst2]{Qin Lou\fnref{1}}
\author[brown]{Xuhui Meng\fnref{1}}
\author[brown,PNNL]{George Em Karniadakis \fnref{2}}

\fntext[1]{The first two authors contributed equally to this work.}
\fntext[2]{Corresponding author: george\_karniadakis@brown.edu (George Em Karniadakis).}
\address[usst1]{School of Energy and Power Engineering, University of Shanghai for Science and Technology, Shanghai 200093, PR China}
\address[usst2]{Shanghai Key Laboratory of Multiphase Flow and Heat Transfer in Power Engineering, University of Shanghai for Science and Technology, Shanghai 200093, PR China}
\address[brown]{Division of Applied Mathematics, Brown University, Providence, RI 02906, USA}
\address[PNNL]{Pacific Northwest National Laboratory, Richland, WA 99354, USA}

\begin{abstract}
The Boltzmann equation with the Bhatnagar-Gross-Krook collision model
(Boltzmann-BGK equation) has been widely employed to describe multiscale flows, i.e., from the hydrodynamic limit to free molecular flow. In this study, we employ physics-informed neural networks (PINNs) to solve forward and inverse problems via the Boltzmann-BGK formulation (PINN-BGK), enabling PINNs to model flows in both the continuum and rarefied regimes. In particular, the PINN-BGK is composed of three sub-networks, i.e., the first for approximating the equilibrium distribution function, the second for approximating the non-equilibrium distribution function, and the third one for encoding the Boltzmann-BGK equation as well as the corresponding boundary/initial conditions. By minimizing the residuals of the governing equations and the mismatch between the predicted and provided boundary/initial conditions, we can approximate the Boltzmann-BGK equation for both continuous and rarefied flows. For forward problems, the PINN-BGK is utilized to solve various benchmark flows given boundary/initial conditions, e.g., Kovasznay flow, Taylor-Green flow, cavity flow, and micro Couette flow for Knudsen number up to 5.  For inverse problems, we focus on rarefied flows in which accurate boundary conditions are difficult to obtain. We employ the PINN-BGK to infer the flow field in the entire computational domain given a limited number of interior scattered measurements on the velocity without using the (unknown) boundary conditions. Results for the two-dimensional micro Couette and micro cavity flows with Knudsen numbers ranging from 0.1 to 10 indicate that the PINN-BGK can infer the velocity field in the entire domain with good accuracy. Finally, we also present some results on using transfer learning to accelerate the training process. Specifically, we can obtain  a three-fold speedup compared to the standard training process (e.g., Adam plus L-BFGS-B) for the two-dimensional flow problems considered in our work.

\end{abstract}


\end{frontmatter}



\section{Introduction}
Multiscale flows spanning the rarefied and continuum regimes are present in diverse applications, such as shale gas flow in porous media \cite{akkutlu2018multiscale, jin2015flow}, vacuum technology \cite{greene2003transitioning,redman2017relevant}, and microfluidics \cite{karniadakis2006microflows,qiao2007modulation,succi2007lattice}. The Boltzmann equation with the Bhatnagar-Gross-Krook collision model (Boltzmann-BGK equation) \cite{bhatnagar1954amodel} is a widely used model for descriptions of both rarefied and continuous flows \cite{chen2007macroscopic,meng2011accuracy}. Several numerical methods have been developed to solve the Boltzmann-BGK equation in recent years since  analytical solutions are difficult to obtain, such as finite difference based methods, e.g., lattice Boltzmann method (LBM) \cite{luo2000some,lallemand2020lattice,luo2000theory,guo2013lattice,succi2001lattice,aidun2010lattice}, finite volume based methods, e.g., gas kinetic scheme (GKS) \cite{lian2000gas,xu2001gas}, the discrete uniform gas kinetic scheme (DUGKS) \cite{guo2013discrete,zhang2019discrete}, and so on.  Successful applications of the aforementioned approaches include multiscale heat transfer \cite{zhang2019implicit}, reactive transport in porous media \cite{meng2020multiscale}, and non-equilibrium flows \cite{xu2011improved},
to name just a few. 


In addition to the conventional numerical methods, deep learning algorithms have emerged recently as an alternative for solving partial differential equations (PDEs), especially in conjunction with sparse data \cite{dissanayake1994neural,sirignano2018dgm,wu2020data,raissi2019physics}. In particular, we focus on the physics-informed neural networks (PINNs) proposed in \cite{raissi2019physics} due to their effectiveness in solving both forward and inverse PDE problems as well as the straightforward implementation. Different from the classical numerical methods in which the partial differential operators need to be discretized, PINNs compute all differential operators in a PDE by using automatic differentiation techniques involved in the backward propagation \cite{raissi2019physics}. Consequently, no mesh is required for the PINN to solve the PDEs, which saves much effort in grid generation \cite{lu2019deepxde,meng2020ppinn}. Another attractive feature is that PINNs are capable of solving the inverse PDE problems effectively \cite{raissi2019physics,meng2020composite} and with the same code that is used for forward problems.  Specifically, PINNs can infer the unknown parameters in a PDE and reconstruct the solution given partial observations on the solution to the PDE. Successful examples of using PINNs for forward and inverse flow problems are: (1) modeling both laminar and turbulent channel flows \cite{jin2020nsfnets},
(2) learning the density, velocity, and pressure fields based on partial observations on the density gradient in high-speed flows \cite{mao2020physics}, and (3) inferring the velocity and pressure fields given measurements from spatio-temporal visualizations of a passive scalar, e.g., temperature or solute concentration \cite{raissi2020hidden}. 

Despite the great progress of PINNs for solving such flow problems, so far only flows in the continuum regime described by the Navier-Stokes or Euler equations have been considered, to the best of our knowledge. However, for flow problems in the rarefied or transitional regimes such as gas flow in shale nanopores \cite{yu2017multiscale,yang2020multiscale}, no existing PINNs or their variants can be applied directly. In addition to PINNs, Han {\sl et al.} proposed to solve the Boltzmann-BGK equation using the deep convolution neural networks \cite{han2019uniformly}, which works well for multiscale flows with a Knudsen number ranging from $10^{-3}$ to 10. However, only forward problems are considered in \cite{han2019uniformly}, i.e., solving the Boltzmann-BGK equation given boundary/initial conditions.  In practical applications, accurate boundary conditions can be quite difficult to obtain, especially for (1) multiscale flows across different flow regimes \cite{bandyopadhyay1987rough,gu2009high,bhattacharya1991nonequilibrium,barisik2012surface,celebi2018molecular}, and (2) rarefied flows within complicated geometries, e.g., porous media \cite{tao2015boundary,singh2017impact,liu2019new,fang2019atomic}.  An interesting question arising here is: can we predict the velocity field when we only have access to a limited number of scattered interior measurements on the velocity rather than the boundary conditions? We refer to this problem as an ``inverse problem" in the present study. To leverage the merits of PINNs for both forward and inverse problems, we propose herein to employ PINNs for solving both the forward and inverse problems via Boltzmann-BGK formulation (PINN-BGK), hence developing a new PINN capability in simulating seamlessly both the continuum and rarefied regimes.

The remainder of the paper is organized as follows. In Section \ref{sec:method}, we describe the details of PINNs for solving the Boltzmann-BGK equation. In Section \ref{sec:forward}, we present the results for modeling some benchmark two-dimensional flows in both the continuum and rarefied regimes. In Section \ref{sec:inverse}, we apply the PINN-BGK to inverse problems, i.e., inferring the flow filed in the entire computational domain given a limited number of scattered measurements on the velocity. A brief summary is then presented in Sec. \ref{sec:summary}.


\section{Methodology}
\label{sec:method}

\subsection{From Boltzmann-BGK equation to discrete velocity Boltzmann-BGK model}
The Boltzmann equation with the simplified collision model, i.e., Bhatnagar-Gross-Krook (BGK) model \cite{bhatnagar1954amodel}, is considered here, which can be expressed as
\begin{align}\label{eq:governingequation}
      \frac{\partial \bm{f}}{\partial t} + \bm{\xi} \cdot \nabla \bm{f}  &= - \frac{1}{\tau} \left(\bm{f} - \bm{f}^{eq} \right)\nonumber\\
      \bm{f}^{neq} &= \left(\bm{f} - \bm{f}^{eq} \right),
\end{align}
where $\bm{f}=\bm{f}(\bm{\xi},\bm{x}, t)$ is the particle distribution function moving with a velocity $\bm{\xi}$ at position $\bm{x}$ and time $t$, $\tau$ is the relaxation time related to the dynamic viscosity $\mu$ and the fluid pressure $p$ as $\tau=\mu/p$,  and $\bm{f}^{eq}$ is the equilibrium distribution  function given by the Maxwellian equilibrium distribution function,
\begin{align}\label{eq:feq}
      \bm{f}^{eq} =  \frac{\rho }{(2 \pi RT)^{D/2}}  \exp \left( -\frac{|\bm{\xi}-\bm{u}|^2}{2 RT} \right), 
\end{align}
where $R$ is the gas constant, $T$ is the temperature, $D$ is the spatial dimension, $\rho$ is the density related to the fluid pressure as $\rho= p/RT$, $\bm{u}$ is the fluid velocity;  $\bm{f}^{neq}$ denotes the non-equilibrium part of the distribution function.

The particle velocity $\bm{\xi}$ is continuous, which makes it computationally inefficient to solve Eq. \eqref{eq:governingequation}. To reduce the computational cost, the discrete velocity model is generally used while keeping the mass and momentum conserved \cite{shan1998discretization,mieussens2000discrete}.  In particular, we can expand $\bm{f}$ in terms of the Hermite polynomials in the velocity space $\bm{\xi}$ \cite{ShanXiaowenJMF2006}:
 \begin{align}
      \bm{f}\left(\bm{x}, \bm{\xi}, t \right) = \omega(\bm{\xi}) \sum^{\infty}_{n = 0} \frac{1}{n!} \bm{a}^{n}(\bm{x}, t) \mathcal{H}^{n} \left(\bm{\xi} \right),
 \end{align}
 where $\bm{a}^{n}$ reads as 
 \begin{align}\label{an}
       \bm{a}^{n}(\bm{x}, t) = \int \bm{f}(\bm{x}, \bm{\xi}, t) \mathcal{H}^{n} (\bm{\xi}) d \bm{\xi}.
 \end{align}
We can then truncate $\bm{f}$ to order $N$, and rewrite the right hand side of Eq. \eqref{an} as
\begin{align}
      \bm{f}^N(\bm{x}, \bm{\xi}, t) \mathcal{H}^{n} (\bm{\xi}) = \omega(\bm{\xi}) \bm{p}(\bm{x}, \bm{\xi}, t), 
\end{align}
where  $\bm{p}$  is a polynomial in $\bm{\xi}$ of a degree not greater than $2N$.
By defining $f_i(\bm{x},t)=\omega_if(\bm{x},\bm{\xi},t)/\omega(\bm{\xi_i})$ for $i=0,1,\cdots,Q-1$, the discrete velocity Boltzmann-BGK model (DVB) can be obtained as follows,
\begin{align}\label{eq:bgk}
      \frac{\partial f_i}{\partial t} + \bm{\xi_i} \cdot \nabla f_i  = - \frac{1}{\tau} \left(f_i - f_i^{eq} \right),
\end{align}
where $f_i^{eq}$ is defined as 
\begin{equation}\label{eq:feq_d}
f_i^{eq}(\bm{x},t)=w_i\rho\left[1+\frac{\bm{\xi}_i\cdot\bm{u}}{RT}+
\frac{(\bm{\xi}_i\cdot\bm{u})^2}{(2RT)^2}-
\frac{\bm{u}\cdot\bm{u}}{2RT}\right],
\end{equation}
to satisfy the mass and momentum conservation laws:
\begin{equation}\label{rhou}
\rho=\sum_i f_i = \sum_i f^{eq}_i,\quad \rho\bm{u}=\sum_i \bm{c}_i f_i = \sum_i \bm{c}_i f^{eq}_i.
\end{equation}
Note that different discrete velocity models should be used for flows with different Knudsen numbers.  More details on this topic can be found in \cite{ShanXiaowenJMF2006}.

\begin{figure}[h]
\centerline{\includegraphics[width=0.7\textwidth]{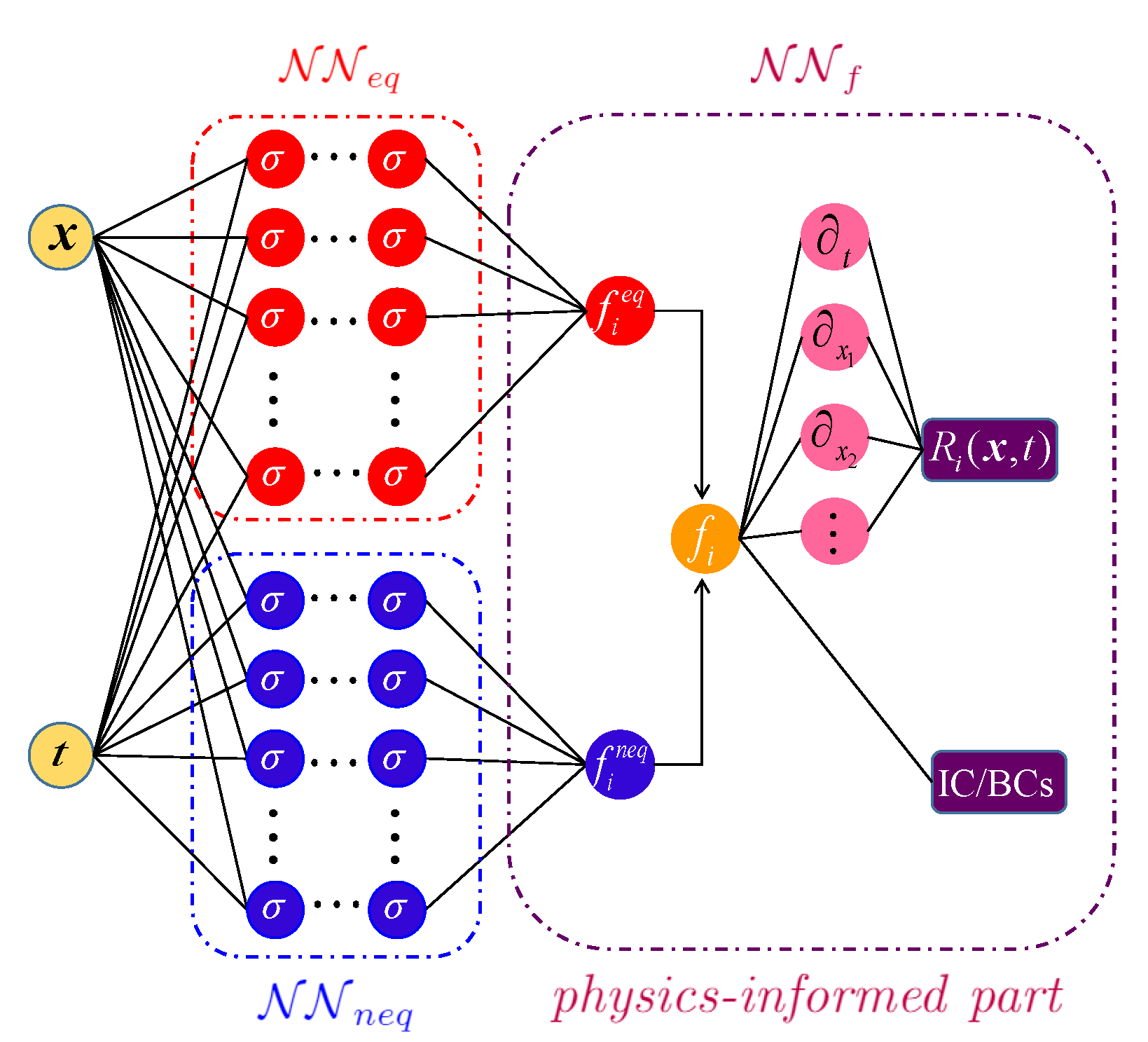}}
\caption{\label{fig:net}
Schematic of PINNs for solving the discrete velocity Boltzmann-BGK equation (PINN-BGK).  $\mathcal{NN}_{eq}$ and $\mathcal{NN}_{neq}$ are employed to approximate the equilibrium ($f^{eq}_i$) and non-equilibrium ($f^{neq}_i$) parts  of the distribution functions, respectively. $\sigma$ is the activation function, which is set as the hyperbolic tangent function in this work. $R_i(\bm{x},t) = \partial_t f_i + \bm{\xi}_i \cdot \nabla f_i  + \left(f_i - f_i^{eq} \right)/\tau$ is the residual of the Boltzmann-BGK equation. IC/BCs represent the initial and boundary conditions. 
}
\end{figure}

\subsection{Physics-informed neural networks for solving DVB model}
\label{sec:pinn}
In this section,  we will extend PINNs for solving both forward and inverse problems via the Boltzmann-BGK formulation. As shown in Fig. \ref{fig:net}, the PINN-BGK is composed of three sub-networks, i.e., one to approximate the equilibrium distribution function, a second one for the non-equilibrium distribution function, and the last one to encode the Boltzmann-BGK equation as well as the corresponding boundary/initial conditions.

Note that the distribution function $f_i(\bm{x},t)$ is decomposed into an equilibrium part (i.e., $f_i^{eq}$) and a non-equilibrium part (i.e., $f_i^{neq}$), i.e., $
f_i(\bm{x}, t) = f_i^{eq} + f_i^{neq}$, in the PINN-BGK. Generally, the magnitudes of $f_i^{eq}$ and $f_i^{neq}$ can be quite different. For continuous flows, we can conduct the following multiscale analysis for Eq. \eqref{eq:bgk} since the relaxation time $\tau$  is a small parameter (e.g., $\tau < 10^{-2}$)  \cite{succi2001lattice,guo2000lattice}.
Specifically, we can expand $f_i$ in terms of $\tau$ as \cite{chen2015comparative} 
\begin{align}\label{eq:msf}
    f_i = f^{(0)}_i + \tau f^{(1)}_i + \tau^2 f^{(2)}_{i} + ... .
\end{align}
Substituting Eq. \eqref{eq:msf} into Eq. \eqref{eq:bgk}, we can obtain the following equations \cite{chen2015comparative}:
\begin{align}
    &\frac{1}{\tau}:~ f^{(0)}_i = f^{eq}_i, \label{eq:f0}\\
    &\tau^0:~ f^{1}_i = -\left(\partial_{t} f^{(0)}_i + \bm{\xi}_i \cdot \partial_{x} f^{(0)}_i \right). \label{eq:f1}
\end{align}
As we truncate $f_i$ to $O(\tau^2)$, the following equation can then be obtained
\begin{align}\label{eq:f_2}
    f_i = f^{(eq)}_i +  \tau f^{(1)}_i + O(\tau^2).
\end{align}
With the aid of Eq. \eqref{eq:f1}, we can rewrite Eq. \eqref{eq:f_2} as
\begin{align}
    f_i \approx f^{eq} - \tau \left(\partial_t f^{(0)}_i + \bm{\xi}_i \cdot \partial_x f^{(0)}_i \right),
\end{align}
which can be further rewritten as
\begin{align}\label{eq:fneq}
    f^{neq}_i = f_i - f^{eq}_i \approx -\tau \left(\partial_t f^{eq}_i + \bm{\xi}_i \cdot \partial_x f^{eq}_i \right),
\end{align}
where $f^{neq}_i$ is the same as in Fig. \ref{fig:net}.
As we can see in Eqs. \eqref{eq:f0} and \eqref{eq:fneq}, $f_i^{eq}$ can be more than 100 times larger than the $f_i^{neq}$ in continuous flows \cite{meng2016localized,guo2002lattice}. If we approximate $f_i$ with one deep neural network (DNN) directly, it is probable that $f_i^{neq}$ will not be resolved accurately since $f_i^{eq}$ dominates over $f_i$. However, $f_i^{neq}$ has a strong influence on the convergence of the governing equation in Eq. \eqref{eq:bgk} due to the fact that $\tau$ is also a small parameter, e.g., $\tau \thicksim O(10^{-4})$ in Sec. \ref{sec:kov} (Numerical results for the effect of $f^{neq}_i$ on predictive accuracy are presented in Sec. \ref{sec:kov}). We therefore apply two DNNs to approximate $f_i^{eq}$ and $f_i^{neq}$, respectively.

In forward problems, we assume that both the initial/boundary conditions and the governing equations are known, and hence we can  solve the Boltzmann-BGK equation by minimizing the following loss function,
\begin{equation}\label{eq:lossfunction_forward}
\begin{split}
&L = L_{Eq} + L_{IC} +  L_{BC} ,\\
&L_{Eq }= \frac{1}{N_e}\sum_{i=0}^{Q-1}\sum_{j=1}^{N_e} |R_i(\bm{x}_j, t_j)|^2,\\
&L_{IC} =  L_{f_{IC}} + L_{m_{IC}},\\
&L_{BC} =  L_{f_{BC}} +  L_{m_{BC}},\\
&L_{f_{IC}} = \frac{1}{N_i}\sum_{i=0}^{Q-1}\sum_{j=1}^{N_i} |f_i(\bm{x}_j, 0) - f_i^*(\bm{x}_j, 0)|^2,\\
&L_{m_{IC}} = \frac{1}{N_i}\sum_{n=1}^{N_i} |\phi(\bm{x}_j, 0) - {\phi^*}(\bm{x}_j, 0)|^2,\\
&L_{f_{BC}} = \frac{1}{N_b}\sum_{i=0}^{Q-1}\sum_{j=1}^{N_b} [ |f_i(\bm{x}_{b_j}, t) - {f_i^*}(\bm{x}_{b_j}, t)|^2 + \\& \quad\quad\quad|\sum_if_i(\bm{x}_{b_j}, t) - \rho^*(\bm{x}_{b_j}, t)|^2 + \\& \quad\quad\quad|\sum_i\bm{\xi}_if_i(\bm{x}_{b_j}, t) - (\rho\bm{u})^*(\bm{x}_{b_j}, t)|^2 ], \\
 &L_{m_{BC}}= \frac{1}{N_b}\sum_{j=1}^{N_b} |\phi(\bm{x}_{b_j}, t) - \phi^*(\bm{x}_{b_j}, t)|^2,
 \end{split}
\end{equation}
where $R_i(\bm{x}, t)$ is residual of the corresponding equation that is defined as
\begin{align}
    R_i: = \frac{\partial f_i}{\partial t} + \bm{\xi}_i \cdot \nabla f_i  + \frac{1}{\tau} \left(f_i - f_i^{eq} \right).
\end{align}
In addition, $L_{Eq}$, $L_{f_{IC}}$, $L_{m{IC}}$, $L_{f_{BC}}$, and $L_{m_{BC}}$ represent the loss function for the residual of the DVB equation, the initial condition of distribution function, the initial condition of the macroscopic variables (i.e., $\rho/u/v$), and the boundary condition of the distribution function, the boundary condition of the macroscopic variables (i.e., $\rho/u/v$). $N_e$, $N_i$, and $N_b$ are the number of the training data for governing equations, initial condition, and boundary conditions, respectively; $\phi$ denotes the macroscopic variables, i.e., $\rho/u/v$; $\bm{x}_b$ represents the locations on the boundary, while $f^*/\phi^*$ denote the exact solutions or measurements.

For inverse problems, we assume that we know the equations as well as a small set of interior measurements on the velocity, but the initial/boundary conditions are unknown. We then infer the velocity field by minimizing the following  loss function:
\begin{equation}\label{eq:lossfunction_inverse}
\begin{split}
&L =  L_{Eq} +  L_{u},\\
&L_{u} = \frac{1}{N_u}\sum_{j=1}^{N_u}|u(\bm{x}_{u_j}, t_{u_j}) - u^*(\bm{x}_{u_j}, t_{u_j})|^2,
\end{split}
\end{equation}
where $L_{Eq}$ is  the same as in Eq. \eqref{eq:lossfunction_forward}, and $L_u$ denotes the mismatch between the predictions and the measurements, $(\bm{x}_u, t_u)$ are the locations of measurements, and $N_e$ and $N_u$ are the  numbers of points for evaluating the residuals of the  governing equation and measurements, respectively.


In general, the pressure or velocity at a boundary is provided for the boundary condition in a flow problem. To solve the discrete velocity Boltzmann-BGK equation, we need to specify appropriate boundary conditions for the discrete distribution functions based on the given pressure/velocity, i.e., kinetic boundary conditions. In the present study, two different kinetic boundary conditions, i.e., Eq. \eqref{eq:fneq} and the diffuse-reflection boundary condition, are employed for the continuous and rarefied flows, respectively. Specifically, (1) in continuous flows, $f^{eq}_i$ ($\bm{\xi}_i\cdot\bm{n}>0$, $\bm{n}$ is the unit inward-wall vector normal to the boundary) can be obtained using the given velocity based on Eq. \eqref{eq:feq_d}, and Eq. \eqref{eq:fneq} can serve as the constraint for the non-equilibrium distribution function at the boundary; (2) For rarefied flows, the gas-surface interaction is implemented via the diffuse-reflection boundary condition \cite{aristov2020kinetic},
\begin{align}
 &f^{neq}_i(t,\bm{x}_b) = \rho_w f_i^{eq}(1, \bm{u}_w) -f^{eq}_i(\rho_w, \bm{u}_w), \quad (\bm{\xi}_i\cdot\bm{n}>0), 
\end{align}
where $\bm{u}_w$ is the wall velocity, and $\rho_w$ is the density at the wall determined by 
\begin{align}
\rho_w = -\frac{\sum_{\bm{\xi}_i\cdot\bm{n}<0}(\bm{\xi}_i\cdot\bm{n})f_i(t,\bm{x}_b)}{\sum_{\bm{\xi}_i\cdot\bm{n}>0}(\bm{\xi}_i\cdot\bm{n})}.
\end{align}
In both the continuous and rarefied flows, Eq.~\eqref{eq:bgk} is enforced at the boundaries for components with $\bm{\xi}_i\cdot\bm{n}<0$. For initial conditions, we can employ the following approximations: $f_i \approx f^{eq}_i$, and $f^{neq}_i = 0$, which is the same as that used in the lattice Boltzmann method \cite{succi2001lattice,guo2013discrete}. We would like to point out that various kinetic initial/boundary conditions have been developed during the past few decades, and  interested readers can refer to \cite{succi2001lattice,guo2013lattice,guo2013discrete,chen2015comparative} for more details.

\section{Forward problems}
\label{sec:forward}
In this section, we will apply the PINN-BGK to model various two-dimensional flow problems given boundary/initial conditions, i.e., (1) continuous flows including time-independent/dependent cases, e.g., Kovasznay flow and Taylor-Green flow; and (2) rarefied flows, i.e.,  micro-Couette flows with Knudsen numbers ranging from 0.01 to 5.

\subsection{Kovasznay flow}
\label{sec:kov}
We first employ the PINN-BGK to simulate the two-dimensional steady incompressible Kovasznay flow in a two-dimensional rectangular domain with a length $-\frac{1}{2} \le x \le 2$ and height $ -\frac{1}{2} \le y \le \frac{3}{2}$. The exact solutions for the flow considered here  are  \cite{kovasznay1948laminar}:
\begin{equation}\label{eq:exact_kovas}
\begin{split}
   u(x,~y) = u_0[1 -\exp{(\lambda x)}\cos {(2\pi y)}], \\
   v(x,~y) = u_0[\frac{\lambda}{2\pi}\exp({\lambda x}) \sin(2\pi y)], \\
   p(x,~y) = p_0[1- \frac{1}{2}\exp{(2\lambda x)}] + C,
   \end{split}
\end{equation}
where $u_0$ and $p_0$ are reference velocity and reference pressure, respectively, $C$ is a constant, $L$ is half the height of the computational domain, $\bm{u} = (u, v)$ represents the fluid velocity, $p$ is the fluid pressure, and 
\begin{align}
\lambda = \frac{Re}{2}-\sqrt{\frac{Re^2}{4}+4\pi^2}, \quad Re=\frac{ u_0 L}{\nu},
\end{align}
where $\nu$ is the kinematic viscosity. In addition, Dirichlet boundary conditions which can be derived from Eq. \eqref{eq:exact_kovas} are imposed on all boundaries.

 Here we employ  the two-dimensional nine-speed
(D2Q9) model \cite{qian1992lattice}, which is widely used for continuous flows.  Specifically,  D2Q9 is defined as $\bm{\xi}_i=c\bm{e}_i$ in which 
\begin{equation}
\begin{split}
c&=\sqrt{3RT}, ~\bm{e}_0  =(0,~0),\\
\bm{e}_1  =-\bm{e}_3&=(1,~0),~\bm{e}_2=-\bm{e}_4=(0,~1),\\
\bm{e}_5  =-\bm{e}_7&=(1,~1),~\bm{e}_6=-\bm{e}_8=(-1,1).  
\end{split}
\end{equation}
In our simulations, Re $ = 10$, $u_0 = 0.1581$, $p_0 = 0.05$, $RT = 100$, $C = RT$, and $L =1 $. The kinematic viscosity and the relaxation time can then be computed as $\nu=Lu_0/Re=0.0158$ and $\tau=\mu/p =\nu/RT= 1.58 \times 10^{-4}$. In addition,  we randomly sample $17,000$ points for the evaluation of residuals, and $300$ points are used on each boundary to provide  Dirichlet boundary conditions. As for the optimization, we first employ the Adam to train the PINN-BGK until the loss is less than $10^{-3}$, then we switch to the L-BFGS-B until convergence.

We note that $\tau \thicksim O(10^{-4})$ in the present case,  leading to $f^{neq}_i \thicksim O(10^{-4})$ according to Eq. \eqref{eq:fneq}. Considering that the single-precision floating-point format is utilized in our code, such a small value of $f^{neq}_i$ will make it difficult to be approximated by DNNs. We then scale the output of $\mathcal{NN}_{neq}$ as $f^{neq}_{NN,i} = 10,000 \times f^{neq}_i$ to make it of the order $O(1)$, where $f^{neq}_{NN,i}$ is the output of the $\mathcal{NN}_{neq}$, and $f^{neq}_i$ is the predicted non-equilibrium distribution function used in Eq. \eqref{eq:bgk}.

\begin{figure}[H]
\centering
\subfigure[]{\label{fig:Loss_kovas}
\includegraphics[width = 0.3\textwidth]{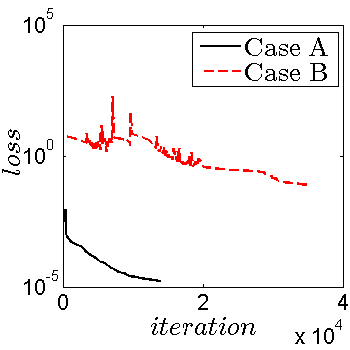}}
\subfigure[]{\label{fig:Eq_err_caseI}
\includegraphics[width = 0.3\textwidth]{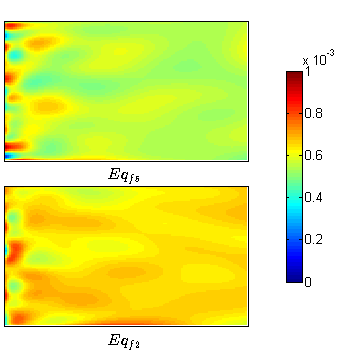}}
\subfigure[]{\label{fig:Eq_err_caseII}
\includegraphics[width = 0.295\textwidth]{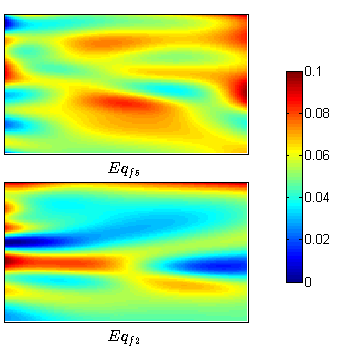}}
\caption{\label{fig:loss_eqerr_kovas} 
PINN-BGK for Kovasznay flows: 
(a) Time histories of the total loss function obtained by different cases. Case A: $f^{neq}_{NN,i} \approx 10000 f^{neq}_i$; Case B: $f^{neq}_{NN,i} \approx  f^{neq}_i$. 
(b) Predicted residuals for $f_2$ and $f_5$ for Case A. 
(c) Predicted  residuals for $f_2$ and $f_5$ for Case B. 
}
\end{figure}

To study the effect of scaling of $f^{neq}_i$ on the predictive accuracy, we compare the results for two different cases, i.e., with and without scaling for $f^{neq}_i$, in Fig. \ref{fig:loss_eqerr_kovas}. In both cases, we employ 5 hidden layers with 40 neurons per layer for $\mathcal{NN}_{eq}$ and $\mathcal{NN}_{neq}$. We see that the training loss for the case with scaling (Case A) is about three orders smaller than that without scaling (Case B). Furthermore, we also present the predicted residuals on a uniform lattice, i.e., $x \times y = 125 \times 100$, for two representative equations, i.e., $f_2$ and $f_5$, in Figs. \ref{fig:Eq_err_caseI}- \ref{fig:Eq_err_caseII}. We observe that the computed residuals for Case B are about two orders larger than those in Case A, which suggests that the scaling of $f^{neq}_i$ can significantly improve the predictive accuracy .

We further compare the relative $L_2$ errors ($err_\psi$) for $u$, $v$ and $f^{neq}_i$ for these two cases, where $err_\psi$ is defined as
\begin{align}\label{eq:error}
err_\psi = \frac{\sqrt{\sum_{x,y}|\psi(x,y,t)-\psi_e(x,y,t)|^2}}{\sqrt{\sum_{x,y}|\psi_e(x,y,t)|^2}},~ \psi = u, ~ v, ~\mbox{and} ~ f_i,
\end{align}
and $\psi_e$ is the exact solution here. In particular, the exact solutions for $u$ and $v$ are in   Eq.~\eqref{eq:exact_kovas},  and the reference solution for $f^{neq}_i$ is computed from Eq.~\eqref{eq:fneq}.  For brevity, we only present the relative errors for $f^{neq}_2$ and $f^{neq}_{5}$ as representatives since the results for the remaining $f^{neq}_i$ are similar as these two.  As displayed in Table ~\ref{table:errorkovas}, the predictions in Case A are all in good agreement with the exact solutions and much better than the results from Case B. These results again indicate that the scaling of $f^{neq}_i$ is able to substantially improve the predictive accuracy.

We note that the errors for $f^{neq}_i$ are about one order greater than those for $u/v$, which may be attributed to the roundoff in the computation of $f^{neq}_i$, i.e., $f^{neq}_i = f^{neq}_{NN,i}/10,000$. Another test case (Case C in Table \ref{table:errorkovas}) is then conducted here, i.e., we replace the $\mathcal{NN}_{eq}$ with the exact solutions for $\rho/u/v$, and only train $\mathcal{NN}_{neq}$ to satisfy the governing equation. Other parameters (e.g., size of $\mathcal{NN}_{eq}$, scaling of $f^{neq}_i$, etc.) as well as the optimization method are all the same as in Case A.  As shown, the errors for $f^{neq}_i$ in Case A are comparable to those in Case C, which validates the present assumption.


\begin{table*}
\caption{
PINN-BGK for Kovasznay flows: Relative $L_2$ errors of velocity ($u,~v$) and two representative non-equilibrium distribution functions. 
Case A: $f^{neq}_{NN,i} \approx 10,000 f^{neq}_i$; 
Case B: $f^{neq}_{NN,i} \approx f^{neq}_i$;  
Case C: $f^{neq}_{NN,i} \approx 10,000 f^{neq}_i$, $\rho$, $u$, and $v$ are from the exact solutions.
}\label{table:errorkovas} \centering
\begin{tabular}{ccccc}
\hline \hline
                &$err_{u}$        &$err_{v}$    &  $err_{f_2^{neq}}$ & $err_{f_5^{neq}}$\\
\hline
Case A         & $0.21\%$        & $0.60\%$               & $3.49\%$          & $4.56\%$ \\
Case B         &$ 65.19\%$       & $ 58.66\%$            &$ 89.41\%$         &$ 82.14\%$\\
Case C        &-          &-               &$1.81\%$           &$2.23\%$ \\
\hline \hline
\end{tabular}
\end{table*}

\begin{table*}
\caption{
PINN-BGK for Kovasznay flows: Relative $L_2$ errors  and computational cost at different Reynolds numbers for different cases. L-BFGS-B (TL): L-BFGS-B optimization with transfer learning. L-BFGS-B only: L-BFGS-B optimization with random initialization. All the computations are performed on a workstation with one Quadro RTX 6000  GPU and Intel(R) Xeon(R) Gold 6242 CPU @ 2.80GHz CPU.  }\label{table:kovas_transferlearning} 
\centering
\begin{tabular}{c|c|c|c|c}
\hline \hline
Re & Case     &$err_u$  &$err_v$ & Walltime (min)   \\ 
\hline
\multirow{3}*{ $20$}
             &L-BFGS-B (TL)      & $0.2\%$ &$0.75\%$     & $20.84$ \\
             
 \cline{2-5} 
             &Adam + L-BFGS-B     & $0.36\%$ &$0.98\%$     & $57.05$ \\
\cline{2-5}

             &L-BFGS-B only     & $0.40\%$ &$1.12\%$    & $32.15$ \\
   
\hline
\multirow{3}*{ $40$}
             &L-BFGS-B (TL)       & $0.22\%$      &$0.89\%$     & $36.14$ \\
            
 \cline{2-5}  
            &Adam + L-BFGS-B  & $0.26\%$      &$0.91\%$     & $63.65$  \\
\cline{2-5}  
          
             &L-BFGS-B only  & $2.63\%$      &$108.78\%$  & $31.47$ \\
\hline
\multirow{3}*{ $60$}
            &L-BFGS-B (TL)       &$0.43\%$       & $1.07\%$     & $28.18$ \\
 \cline{2-5}  
            &Adam + L-BFGS-B  &$0.27\%$   & $1.06\%$    &$75.87$  \\
\cline{2-5} 
          
             &L-BFGS-B only  &$3.08\%$       &$119.43\%$     & $32.76$ \\
\hline
\hline
\end{tabular}
\end{table*}

To accelerate the training process,  we employ the transfer learning strategy, which has been widely used to enhance the convergence of training for diverse deep learning problems \cite{yang2013theory,lu2015transfer}. In particular, we employ the PINN-BGK to model the flows for three different Reynolds numbers, i.e., Re $ = 20$, 40, and 60. The weights/biases for each case are initialized with those for the case Re $=10$ on its completion of training. Note that only the L-BFGS-B algorithm is used for optimizing the loss function in the transfer learning cases. Here, we also present the results for using the Adam plus L-BFGS-B as used in the previous case for comparison. As we can see, comparable accuracy can be obtained by using the L-BFGS-B with transfer learning and the standard training process used in this work, i.e., first Adam and then L-BFGS-B. However, we can obtain up to a three-fold speedup with transfer learning as compared to the  latter case.

It is worth mentioning that the L-BFGS-B algorithm is based on the Newton's method, hence the computational accuracy strongly depends on the initial guess. In other words, it is easy for the L-BFGS-B to converge to a local optimal without good initialization of the weights/biases in DNNs. For demonstration, we then present the results for using the L-BFGS-B with random initialization for the DNNs (which is referred to as L-BFGS-B only in Table \ref{table:kovas_transferlearning}). As shown, the computational time for the L-BFGS-B (TL) and L-BFGS-B only is similar for all the three test cases. However, the L-BFGS-B with transfer learning can achieve much better accuracy than the L-BFGS-B with the random initialization. In particular, the predictive accuracy for the L-BFGS-B (TL) is about two orders better than the L-BFGS-B only for the vertical velocity in the last two cases, i.e., Re $=40$, and 60.

Finally, we note that the multiscale analysis in Sec. \ref{sec:pinn} only holds for continuous flows in which $\tau$ is a small parameter. In the present study, we assume  $f^{neq}_i$ has a similar magnitude as $ O(\tau)$ in all test cases, and scale $f^{neq}_{NN,i}$ to $O(1)$ based on $\tau$ used in the corresponding case to improve the predictive accuracy.


\subsection{Taylor-Green flow}
We proceed to model a time-dependent flow problem, i.e.,  the Taylor-Green flow, using the PINN-BGK. The computational domain is defined as $ x \in [-\pi, \pi], ~y \in [-\pi,\pi]$. The exact solution is set as follows
\begin{equation}\label{eq:taylorgreenexactsolution}
\begin{split}
   u(x,~y,~t) &= -\cos (x) \sin (y) e^{-2t\nu}, \\
   v(x,~y,~t) &= -\sin (x) \cos (y) e^{-2t\nu}, \\
   p(x,~y,~t) &= -\frac{1}{4}[\cos(2x) + \cos(2y)]e^{-4t\nu},
\end{split}
\end{equation}
where $u$, $v$ are the velocity components in the horizontal and vertical directions, respectively, and $p$ denotes the fluid pressure.  Finally, all the initial and boundary conditions are derived from Eq. \eqref{eq:taylorgreenexactsolution}.


 In simulations, we set $RT=100$, $\nu=0.01$,  and $t\in[0,10]$.  As for the training data, $300$ random points are employed on each boundary while   $160,000$ random points in the spatial-temporal domain are used to compute the residuals of Eq. \eqref{eq:bgk}. Both $\mathcal{NN}_{eq}$ and $\mathcal{NN}_{neq}$ contain $6$ hidden layers with $80$ neurons per layer. Similar as  the first test case, two optimization algorithms, i.e., Adam and L-BFGS-B are used. Specifically, we first train the DNNs using the Adam optimizer until the loss is less than $1.0 \times 10^{-4}$  with an initial learning rate of $5 \times 10^{-4}$, and then we switch to the L-BFGS-B  optimization until convergence. 

\begin{figure}[H]
\centering
\subfigure[]{\label{fig:uc_taylora}
\includegraphics[width = 0.45\textwidth]{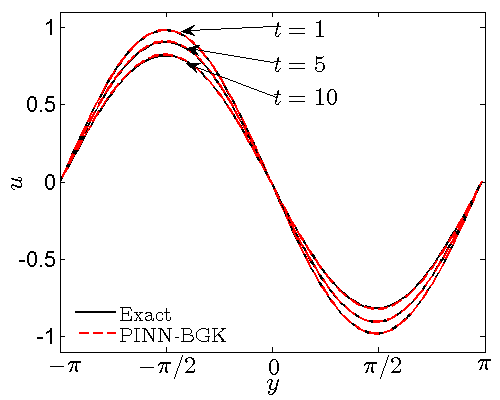}}
\centering
\subfigure[]{\label{fig:vc_taylorb}
\includegraphics[width = 0.45\textwidth]{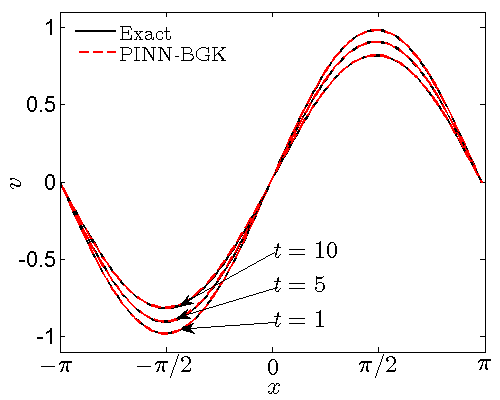}}
\caption{\label{fig:u_taylor} 
PINN-BGK for Taylor Green flows: 
Predicted velocity profiles. (a) $u$ component for $x=0$ at $t = 1$, 5, and 10. 
(b) $v$ component for for $y=0$ at  $t = 1$, 5, and 10. 
}
\end{figure}

 The predicted velocity profiles at three representative times, i.e.,  $t=1$, 5 and 10,  are illustrated in Fig.~\ref{fig:u_taylor}. We see that the predicted velocities agree well with the exact solutions. Furthermore, we present the relative $L_2$ errors for the predicted $u/v/p$ in more detail in Table~\ref{table:taylorerr},  which demonstrates the good accuracy of the PINN-BGK for modeling unsteady flows.

\begin{table*}[h]
\caption{PINNs for Taylor Green flows: Relative $L_2$ errors of pressure ($p$) and velocity ($u,v$)  at different times.}\label{table:taylorerr} \centering
\begin{tabular}{ccccccc}
\hline \hline
Time              &$t=0.0$    &$t=2.0 $   &$t=4.0$    &$t=6.0 $    &$t=8.0$       &$t=10.0$ \\
\hline
$err_{p}$       &$0.00092\%$   &$ 0.0011\%$  & $0.0012\%$   & $0.0015\%$  &$0.002\%$    & $0.0037\%$ \\
$err_{u}$       &$0.0058\%$   &$ 0.16\%$  & $0.24\%$   & $0.30\%$  &$0.35\%$          & $0.48\%$ \\
$err_{v}$       &$0.0051\%$  &$ 0.16\%$  & $0.26\%$   & $0.35\%$  &$0.44\%$          & $0.55\%$ \\
\hline \hline
\end{tabular}
\end{table*}

\subsection{Regularized Cavity flow}
Here we model the cavity flow using the PINN-BGK. The computational domain is set as $x\in[0,~1]$, and $y\in[0,~1]$. The upper boundary is moving from the left to right with a velocity $u_w$ defined as  \begin{equation}\label{eq:uw_cavity}
u_w=u_{max}\left(1-\frac{\cosh[r(x-L/2)]}{\cosh(rL/2)}\right),
\end{equation}
where $r = 10$ is a constant, $u_{max}=1.0$ is the maximum velocity at the top wall, and $L = 1$ is the characteristic length of the cavity. Note that the velocity is continuous at the two upper corners, which can alleviate the effect of singularity on the computational accuracy.

\begin{figure}[H]
\centering
\subfigure[]{\label{fig:cavitya} 
\includegraphics[width = 0.45\textwidth]{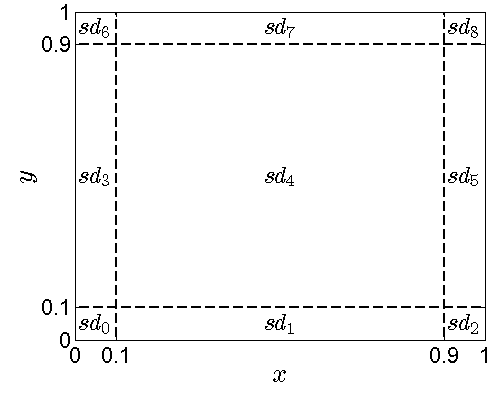}}
\subfigure[]{\label{fig:cavityb} 
\includegraphics[width = 0.45\textwidth]{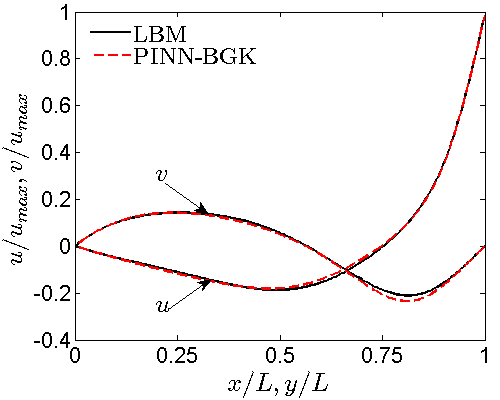}}
\caption{\label{fig:cavity} 
PINN-BGK for cavity flows: 
(a) Schematic of the subdomains for cavity flow. 
(b) Predicted $u$- and $v$-velocity profiles along $y = 0.5$ and $x = 0.5$, respectively.
}
\end{figure}

In simulations, $Re = L u_{max}/\nu = 100$, where $\nu$ is the kinematic viscosity. We  employ  $4$ hidden layers  with $50$ neurons per layer for $\mathcal{NN}_{eq}$, and $8$ hidden layers with $60$ neurons per layer for $\mathcal{NN}_{neq}$. As for the training data, $257$ uniformly distributed points on each boundary are used to provide Dirichlet boundary conditions,  and $45,000$ random points are employed to  compute the residuals of the equations. To accurately predict the boundary layers, we divide the whole computational domain into nine subdomains ($sd_0,~sd_1,\cdots,sd_8$) as shown in Fig.~\ref{fig:cavitya}, which is similar as the nonuniform meshes used in the conventional numerical methods \cite{lee2001characteristic,yang2019improved}.  In each subdomain,  5000 random points are employed to evaluate the residuals of the governing equations.  For the optimization, we first train the DNNs using the Adam optimization with an initial learning rate $1.0 \times 10^{-4}$ for $50,000$ iterations, and then train the DNNs for 50,000 steps using Adam with a smaller learning rate $1.0 \times 10^{-5}$. Finally, we switch to the L-BFGS-B until convergence.

The predicted velocity profiles along the central lines ($x = 0.5$ and $y = 0.5$) are illustrated in Fig.~\ref{fig:cavityb}. In addition, the results from the lattice Boltzmann method (LBM) \cite{guo2000lattice} with $256 \times 256$ uniform grid are also displayed in Fig.~\ref{fig:cavity} as reference. As we can see, the predictions using the PINN-BGK agree well with the results from the LBM.

\subsection{Micro Couette flow}
\label{sec:couette_forward}
The PINN-BGK is now applied to simulate the micro Couette flow between two parallel plates located at $y=-0.5$ and $0.5$. The top and bottom walls  move with a constant velocity $u_w$ in an opposite direction.  For rarefied flows, the Knudsen number $Kn$ is defined as $Kn = \lambda/L$, where $\lambda$ is the mean free path, and $L$ is the characteristic length of the system. Note that $\lambda$ is related to the relaxation time $\tau$ in the Boltzmann-BGK equation as $\lambda =  \tau\sqrt{\pi RT/2}$ \cite{guo2013discrete}. Based on $Kn$, gas flows can be divided into four regimes: the hydrodynamic regime ($Kn<0.001$), the slip regime ($0.001<Kn<0.1$), the transition regime (0.1<$Kn<10$), and the free molecular flow regime ($Kn>10$) \cite{meng2011accuracy}. Here, we employ the PINN-BGK to model the flows in two regimes, i.e., the slip regime ($0.001<Kn<0.1$), and the transition regime ($0.1<Kn<10$). In addition, the two-dimensional-sixteen-velocity (i.e., D2Q16) model is employed for the slip regime while a two-dimensional $28\times28$ velocity model ($28\times28$-DVB) is used for the flows in transition regime.  Interested readers can refer to \cite{ShanXiaowenJMF2006,galant1969gauss} for more details on the two discrete velocity models.


\begin{figure}[H]
\centering
\subfigure[$Kn=0.01$]{\label{fig:uplot_mmicorcouette_kn0.01}
\includegraphics[width = 0.3\textwidth]{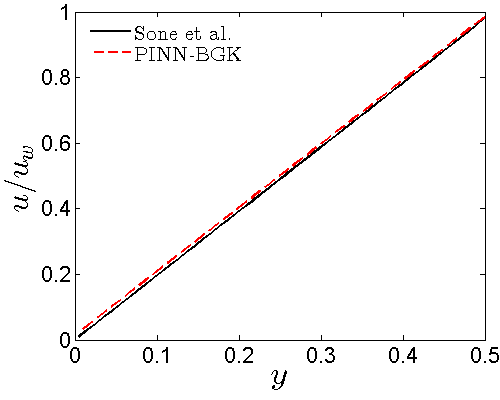}}
\subfigure[$Kn=0.05$]{\label{fig:uplot_mmicorcouette_kn0.05}
\includegraphics[width = 0.3\textwidth]{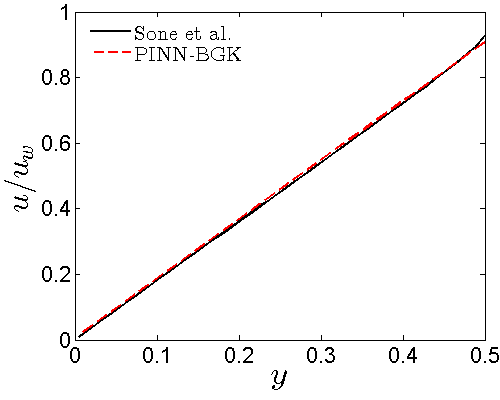}}
\subfigure[$Kn=0.09$]{\label{fig:uplot_mmicorcouette_kn0.08}
\includegraphics[width = 0.3\textwidth]{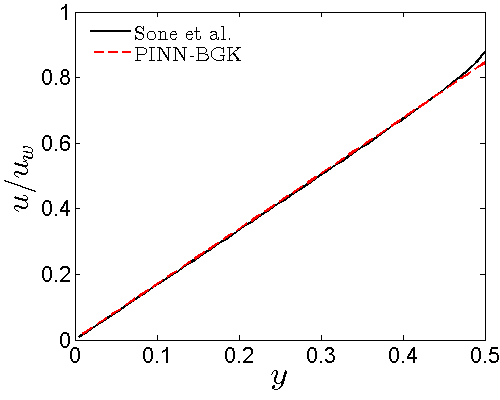}}
\caption{\label{fig:u_slip_couette} 
PINN-BGK for micro-Couette flows: Predicted horizontal velocity profiles in slip regime using the D2Q16 model. 
}
\end{figure}

We first apply the PINN-BGK to simulate  the Couette flow in the slip regime. In particular, three different cases, i.e., $Kn=0.01,~0.05$, and $0.09$, are considered.  For all the test cases, we set $u_w=0.1$, $RT=1.0$. In addition,  both $\mathcal{NN}_{eq}$ and $\mathcal{NN}_{neq}$ have $6$ hidden layers  with $30$ neurons per layer.  In our simulations, $100$ uniformly distributed points are used on each boundary to provide Dirichlet boundary conditions,  and $10,000$ random points are employed to compute the residuals of the equations. Similarly, both the Adam and L-BFGS-B are used for optimization as follows: the Adam optimizer with an initial learning rate $10^{-3}$ is first employed until the loss is less than $10^{-3}$, and then L-BFGS-B is applied until convergence.

\begin{figure}[H]
\centering
\subfigure[]{\label{fig:couettesche}
\includegraphics[width = 0.41\textwidth]{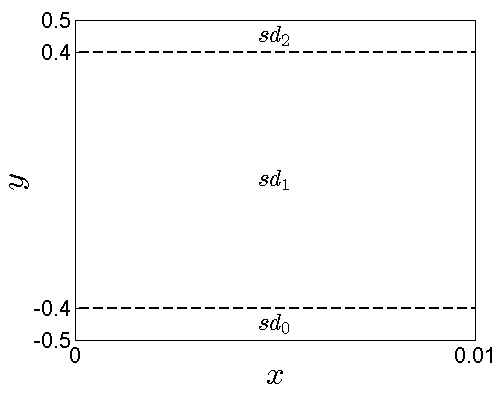}}
\subfigure[]{\label{fig:uc_couette_tran}
\includegraphics[width = 0.4\textwidth]{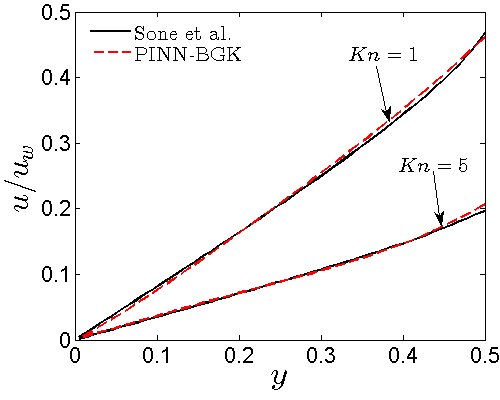}}
\caption{\label{fig:couette_tran} 
PINN-BGK for micro-Couette flows in transition regime: 
(a) Schematic of subdomains. 
(b) Predicted horizontal velocity profiles. 
}
\end{figure}

The horizontal velocity profiles predicted by PINN-BGK are displayed in Fig.~\ref{fig:u_slip_couette}. In addition, the solutions to the linearized Boltzmann equation \cite{Sone1990} are presented as reference. We observe  velocity slip at the upper boundary in all test cases, which is well captured by the PINN-BGK. Specifically,  the relative $L_2$ errors between the PINN-BGK and the reference solutions are $2.1\%,~1.12\%$, and $1.32\%$ for $Kn = 0.01,~0.05$, and $0.09$, respectively, suggesting that the PINN-BGK is capable of providing accurate predictions for the flows in the slip regime.

We now simulate the Couette flow in the transition regime. Here two different cases, i.e.,  $Kn = 1$ and 5 are considered. In both cases, $\mathcal{NN}_{eq}$ has 2 hidden layers with $30$ neurons per layer, and  $\mathcal{NN}_{neq}$ has $2$ hidden layers with $40$ neurons per layer. Here the $28\times 28$ discrete velocity model is employed, in which $RT$ is set to be $1/2$.  Note that the model is more computationally expensive than the previous cases, so we utilize the minibatch approach with a batch size of 800 for both cases to accelerate the training. In addition, $100$ uniformly distributed points on each boundary are employed to enforce the boundary condition. To accurately capture the slip boundary layers near the upper and bottom walls, we divide the computational domain into $3$ subdomains as shown in Fig.~\ref{fig:couettesche}. In $sd_0$ and $sd_2$, $800$ uniformly distributed residual points are employed, while  $1,600$ uniformly distributed points are employed to compute the residuals of equations in $sd_1$.  The Adam optimizer with an initial learning rate $10^{-3}$ is used for optimization.  The predicted horizontal velocity profiles from the PINN-BGK for both cases are depicted in Fig.~\ref{fig:uc_couette_tran}; we observe that they agree well with the reference solutions from the linearized Boltzmann equation in \cite{Sone1990}.

\section{Inverse problems}
\label{sec:inverse}

In this section, we test the performance of the PINN-BGK for inverse problems. Specifically, we focus on rarefied flows with unknown velocity boundary conditions. We assume that we only have partial measurements on the velocity inside the computational domain. The objective is to infer the entire flow field based on a limited number of scattered observations on velocity with the aid of the   Boltzmann-BGK equation.  

\subsection{Micro Couette flows}



\begin{figure}[h]
\centering
\subfigure[]{\label{fig:positions_couette_inverse}
\includegraphics[width = 0.45\textwidth]{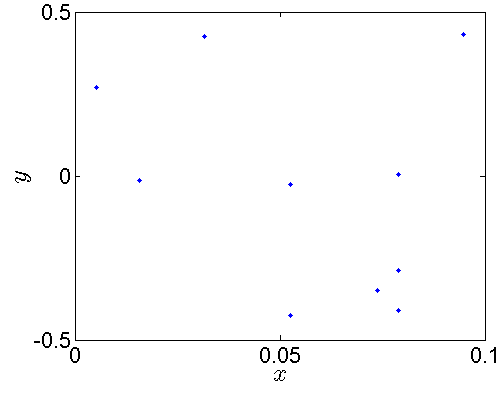}
}
\subfigure[]{\label{fig:u_couette_inverse}
\includegraphics[width = 0.45\textwidth]{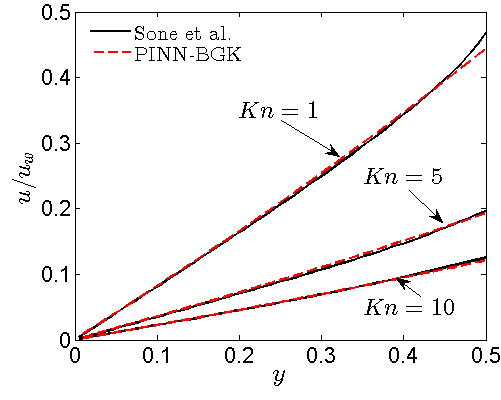}
}
\caption{\label{fig:couette_inverse} 
PINN-BGK for inverse problems in micro Couette flow: 
(a) Locations of the sensors for velocity. 
(b) Predicted velocity profiles along the vertical central lines for different Knudsen numbers.
}
\end{figure}

We first use the PINN-BGK to infer the velocity field for the micro Couette flow based on a small set of  measurements on velocity inside the computational domain.
The computational domain is set as $0\leq x\leq 0.1$, $-0.5\leq y \leq 0.5$. The top and bottom walls move with a constant velocity (i.e., $u_w = 0.1$) in an opposite direction, which is the same as the case in Section \ref{sec:couette_forward}.  Three different cases, i.e., $Kn = 1$, 5, and 10, are considered. We therefore employ the $28\times28$-discrete velocity model as in Section  \ref{sec:couette_forward} \cite{ShanXiaowenJMF2006,galant1969gauss}.  For all the test cases, $RT$ is set to be $1/2$. We then assume that we have 10 sensors for the velocity, which are randomly located  inside the computational domain, as shown in Fig.~\ref{fig:positions_couette_inverse}. In addition, we utilize $4$ hidden layers with $30$ neurons per layer for both $\mathcal{NN}_{eq}$ and $\mathcal{NN}_{neq}$, and $5,000$ ($N_e = 5,000$) residual points are randomly selected inside the computational domain. 
To train the NNs, the Adam optimizer with an initial learning rate $10^{-3}$ is first used until the loss is less than $10^{-4}$, and then the LBFGS-B optimizer is employed until convergence. 

The predicted velocity profiles using the PINN-BGK are displayed in Fig.~\ref{fig:u_couette_inverse}, and the solutions from  the linearized Boltzmann equation \cite{Sone1990} are presented as reference. As shown, the predictions from the PINN-BGK agree well with the reference solutions for all cases. Furthermore, the relative $L_2$ errors between the PINN-BGK and the reference are $2.04\%,~2.04\%$, and $1.44\%$ for $Kn=1,~5$, and $Kn=10$, respectively, which demonstrates that the PINN-BGK is capable of inferring the velocity field based on partial interior scattered measurements on velocity without the explicit knowledge of boundary conditions.

\begin{figure}[H]
\centering
\includegraphics[width = 0.5\textwidth]{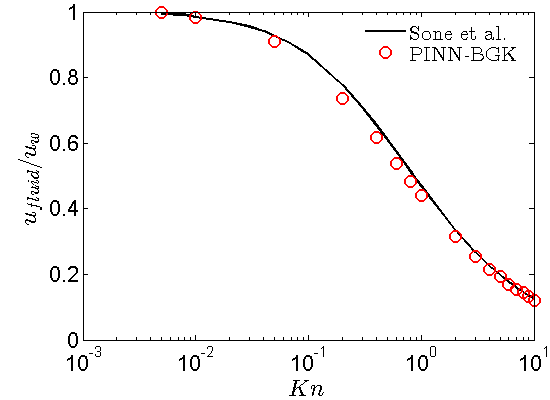}
\caption{\label{fig:u_kn_couette} 
PINN-BGK for inverse problems in micro Couette flow: normalized velocity at the upper boundary for different $Kn$. $u_w$: moving velocity of upper wall.
}
\end{figure}

In addition to estimate the velocity field based on partial interior observations on the velocity, we can also employ the PINN-BGK to predict the slip velocity at the boundary for different Knudsen numbers.  Here, we use the PINN-BGK to infer the fluid velocity at the upper boundary  for $Kn$ ranging from 0.01 to 10. As shown in Fig. \ref{fig:u_kn_couette}, the predicted velocities  are in good agreement with the reference solutions in \cite{Sone1990}, which clearly demonstrates the capability of the PINN-BGK for modeling multiscale flows. Also, we would like to point out that the predicted velocities at the boundary can serve as boundary conditions for other numerical methods, which need explicit boundary conditions to solve the Boltzmann-BGK equations, such as the discrete unified gas kinetic scheme \cite{guo2013discrete}, the unified gas kinetic scheme \cite{chen2017unified}, etc.

\subsection{Micro cavity flows}

We now employ the PINN-BGK to infer the velocity field for the micro cavity flow based on partial interior measurements on velocity, in which the computational domain is set as $0\leq x\leq 1$,  $0 \leq y \leq 1$. The top wall of the cavity moves along the $x$ direction with the velocity defined in  Eq.~\eqref{eq:uw_cavity} in which $u_{max}=0.15$ and $r=10$ while the other walls are stationary. Again, we assume that only a few sensors for the fluid velocity inside the  domain are available while the boundary conditions are unknown.

\begin{figure}[H]
\centering
\subfigure[]{\label{fig:position_cavity_inverse}
\includegraphics[width = 0.45\textwidth]{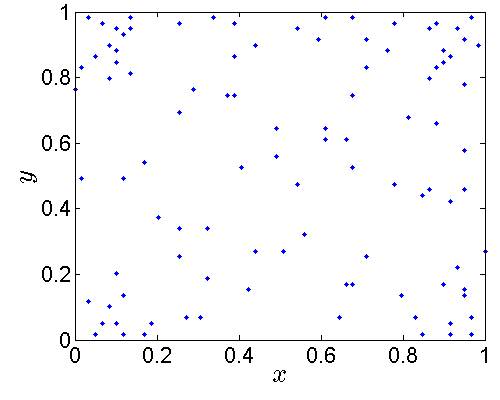}
}
\subfigure[]{\label{fig:kn01_cavity_inverse}
\includegraphics[width = 0.45\textwidth]{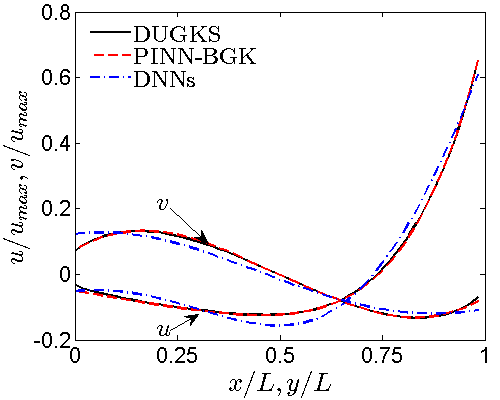}
}
\subfigure[]{\label{fig:kn1_cavty_inverse}
\includegraphics[width = 0.45\textwidth]{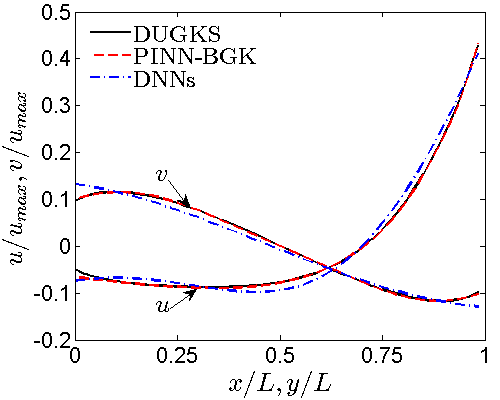}
}
\subfigure[]{\label{fig:kn10_cavity_inverse}
\includegraphics[width = 0.45\textwidth]{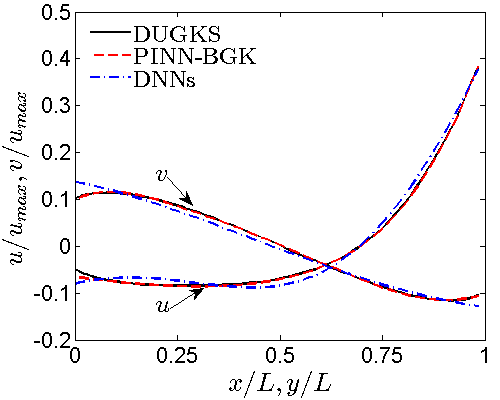}
}
\caption{\label{fig:cavity_inverse} 
PINN-BGK for inverse problems in micro cavity flows with random training data: 
(a) Locations of sensors for $\bm{u}$; 
(b - d): Predicted velocity profiles across the cavity center for (b) $Kn=0.1$, (c)  $Kn=1.0$, (d)  $Kn=10$.
}
\end{figure}

Here we test three different cases, i.e., $Kn = 0.1$, 1, and 10. In addition, the discrete unified gas kinetic scheme (DUGKS) \cite{guo2013discrete}, which is a finite-volume solver for the Boltzmann-BGK equation, is used to generate training data as well as provide reference solutions. In the DUGKS simulations, the computational domain is discretized into  $60 \times 60$ uniform grids.  Similar as in \cite{guo2013discrete,chen2017unified}, the Newton-Cotes quadrature is used to discrete the velocity space, and $81 \times 81$ nodes distributed uniformly in $[-4\sqrt{RT},4\sqrt{RT}]\times[-4\sqrt{RT},4\sqrt{RT}]$ are employed. $RT$  is set to be $1/2$, $\nu$ and $\tau$ can then be computed using $Kn$ and $RT$. In the PINN-BGK, we employ the same discrete velocity model as in DUGKS. In addition, we first assume that we have 100 sensors for the velocity, which are randomly distributed inside the computational domain, see   Fig.~\ref{fig:position_cavity_inverse}.  Two hidden layers with $30$ neurons per layer are used for $\mathcal{NN}_{eq}$ while six hidden layers with $30$ neurons per layer are employed for $\mathcal{NN}_{neq}$.  Also, $12,000$ random points are employed to compute the residuals for the equation in the loss function. Similarly, the minibatch  approach with a batch size of 128 is employed to accelerate the training.

The predicted velocity profiles from the PINN-BGK across the cavity center for different Knudsen numbers are presented in
Figs.~\ref{fig:kn01_cavity_inverse}-\ref{fig:kn10_cavity_inverse}. In addition, the results from  DUGKS  are also presented as reference solutions in Figs.~\ref{fig:kn01_cavity_inverse}-\ref{fig:kn10_cavity_inverse}. We can see that  the PINN-BGK predictions are in good agreement with the results from the DUGKS at the three Knudsen numbers, demonstrating the capability of the PINN-BGK to model multiscale transitional flows with unknown  velocity boundary conditions.

To further demonstrate the effectiveness of the PINN-BGK, we also present the results for regression (denoted as DNNs in Figs.~\ref{fig:kn01_cavity_inverse}-\ref{fig:kn10_cavity_inverse}), i.e., we employ $\mathcal{NN}_{eq}$ to predict the velocity profiles based on the same measurements on velocity without the constraint of the governing equation. We observe that the predictions from the PINN-BGK are better than those from DNN regression, especially for areas near the boundaries. We further compute the relative $L_2$ errors for the PINN-BGK and regression, in which the results from the DUGKS are used as reference solutions. As shown in Table~\ref{table:errorcavity_inverse}, 
the errors for the PINN-BGK are less than $5\%$ in all cases, suggesting that the PINN-BGK can accurately reconstruct the velocity field even without the use of boundary conditions. In addition, the errors for regression are larger than $15 \%$ in 
all cases, which means that enforcing the constraint of the governing equations in the loss function helps greatly to improve the accuracy of the predictions.

\begin{table*}
\caption{
PINN-BGK for inverse problems in micro cavity flows  with random training data: Relative $L_2$ relative errors for velocity ($u,~v$) at different Knudsen numbers. DNNs: results from regression without the constraint of PDEs.  
Reference solutions: DUGKS with $60 \times 60$ uniform grids. 
}\label{table:errorcavity_inverse} \centering
\begin{tabular}{c|c|c|c|c}
\hline \hline\multirow{2}*{~$Kn$~}

             &\multicolumn{2}{c|}{$err_u$}     &    \multicolumn{2}{c}{$err_v$} \\
\cline{2-5} 

             & PINN-BGK    & DNNs   & PINN-BGK    & DNNs \\
\hline

~$0.1$~      &$ 3.36\%$  &  $ 18.68\%$     & $4.27\%$      & $ 19.65\%$      \\
~$1$ ~       &$3.39\%$   &   $ 16.02\%$      & $3.36\%$      &$ 15.71\%$    \\
~$10$~       &$3.66\%$   &   $ 15.13\%$     &$2.99\%$        &   $ 18.81\%$  \\
\hline \hline
\end{tabular}
\end{table*}

\begin{figure}[H]
\centering
\subfigure[]{\label{fig:xf_lines}
\includegraphics[width = 0.45\textwidth]{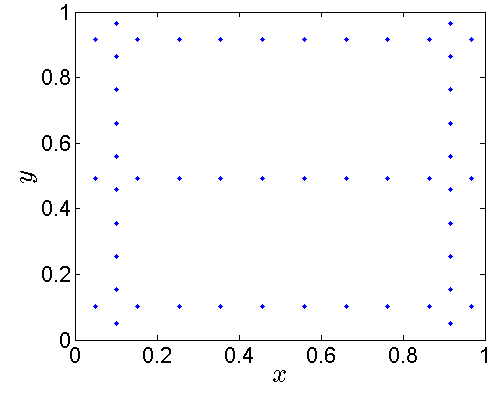}
}
\subfigure[]{\label{fig:uc_kn01_lines}
\includegraphics[width = 0.45\textwidth]{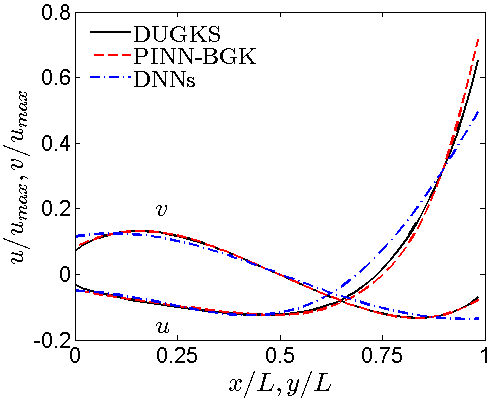}
}
\subfigure[]{\label{fig:uc_kn1_lines}
\includegraphics[width = 0.45\textwidth]{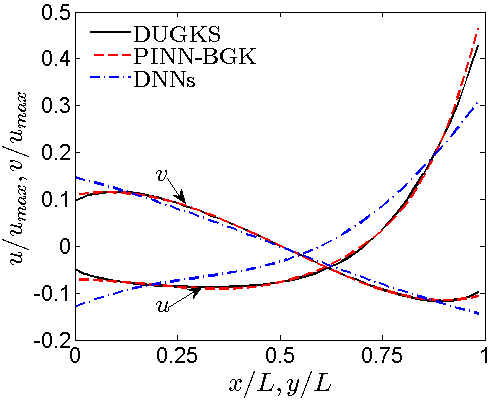}
}
\subfigure[]{\label{fig:uc_kn10_lines}
\includegraphics[width = 0.45\textwidth]{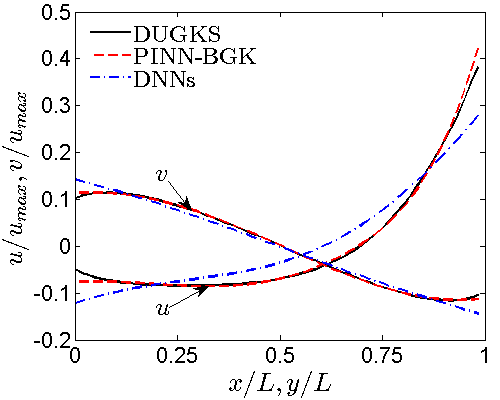}
}
\caption{ 
PINN-BGK for inverse problems in micro cavity flows with uniform training data (Case I): 
(a) Locations of sensors for $\bm{u}$. The 50 training samples are uniformly distributed along 5 lines.  
(b - d): Predicted velocity profiles across the cavity center for (b) $Kn=0.1$, (c)  $Kn=1.0$, (d)  $Kn=10$.
}\label{fig:uc_cavity_lines}
\end{figure}

\begin{figure}[H]
\centering
\subfigure[]{\label{fig:xf_caseI}
\includegraphics[width = 0.45\textwidth]{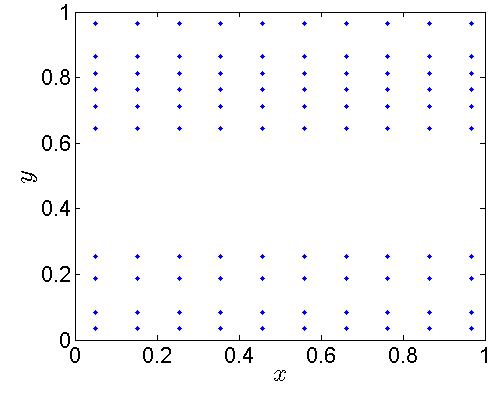}
}
\subfigure[]{\label{fig:uc_kn01_caseI}
\includegraphics[width = 0.45\textwidth]{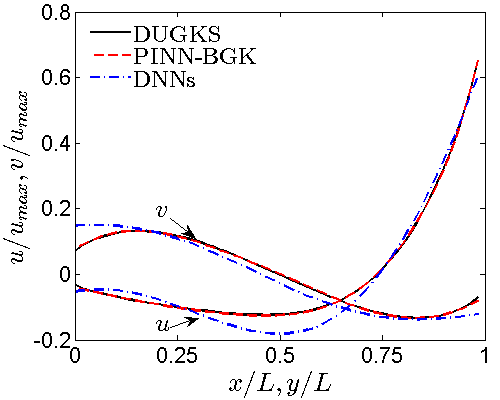}
}
\subfigure[]{\label{fig:uc_kn1_caseI}
\includegraphics[width = 0.45\textwidth]{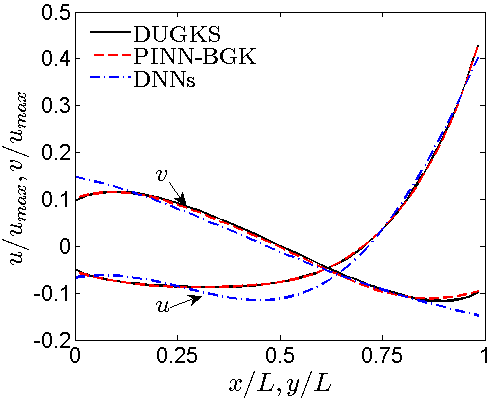}
}
\subfigure[]{\label{fig:uc_kn10_caseI}
\includegraphics[width = 0.45\textwidth]{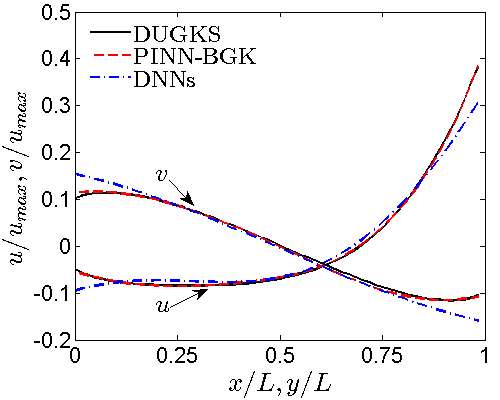}
}
\caption{ 
PINN-BGK for inverse problems in micro cavity flows with uniform training data (Case II): 
(a) Locations of sensors for $\bm{u}$. The 100 training samples are uniformly distributed along 10 lines.  
(b - d): Predicted velocity profiles across the cavity center for (b) $Kn=0.1$, (c)  $Kn=1.0$, (d)  $Kn=10$.
}\label{fig:uc_cavity_CaseI}
\end{figure}

\begin{table*}
\caption{
PINN-BGK for inverse problems in micro cavity flows with uniformly sampled training data: Relative $L_2$ errors for velocity ($u,~v$) at different Knudsen numbers. DNNs: results from regression without the constraint of PDEs.  
Reference solutions: DUGKS with $60 \times 60$ uniform grids. Case I: 50 measurements on $\bm{u}$ are uniformly distributed along 5 lines. Case II: 100 measurements on $\bm{u}$ are uniformly distributed along 10 lines.
}\label{table:relative_errorcavity_inverse_uniformgrid} \centering
\begin{tabular}{c|c|c|c|c|c}
\hline \hline\multirow{2}*{~}

  \multirow{2}*{Case}     & \multirow{2}*{~$Kn$~}  &\multicolumn{2}{c|}{$err_u$}     &    \multicolumn{2}{c}{$err_v$} \\
\cline{3-6} 

       &      & PINN-BGK    & DNNs   & PINN-BGK    & DNNs \\
\hline\multirow{3}*{Case I}

&$0.1$      &$ 9.37\%$  &  $ 25.45\%$     & $36.58\%$      & $21.52\%$      \\
&$1$        &$8.51\%$   &   $ 35.29\%$      & $21.39\%$      &$ 23.07\%$    \\
&$10$       &$8.56\%$   &   $ 31.87\%$     &$17.46\%$       &$ 20.62\%$  \\
\hline\multirow{3}*{Case II}
&$0.1$      &$ 3.26\%$  &  $ 20.22\%$     & $5.0\%$      & $ 19.47\%$      \\
&$1$        &$3.5\%$   &   $ 17.99\%$      & $4.49\%$    &$ 15.99\%$    \\
&$10$       &$3.62\%$   &   $ 29.10\%$     &$4.04\%$     & $ 21.69\%$  \\
\hline \hline
\end{tabular}
\end{table*}

Due to the fact that the measurements at arbitrary locations may be difficult to obtain in practical applications, next we first test the case in which we assume that 50 measurements on $\bm{u}$ are uniformly distributed along 5 lines (Fig.~\ref{fig:xf_lines}, Case I). 
In simulations of the present case, the parameters (e.g., the architectures of DNNs, optimizer, etc.) are kept the same as those in the previous one. The predicted velocity profiles for $Kn=0.1,~1$, and $10$ across the cavity center are presented in Fig.~\ref{fig:uc_cavity_lines}. As observed, the results predicted by the PINN-BGK here are not as accurate as those in Fig. \ref{fig:cavity_inverse}, especially for areas near the boundaries. Specifically, the relative $L_2$ errors for velocity (Case I in Table \ref{table:relative_errorcavity_inverse_uniformgrid}) are much larger than the results from PINN-BGK in Table \ref{table:errorcavity_inverse}. We then increase the number of training samples to 100, which are uniformly distributed along 10 lines (Fig.~\ref{fig:xf_caseI}, Case II). Specifically, more measurements are added near the vortex zone in the present case. Similarly, we employ the same parameters (e.g., the architectures of DNNs, optimizer, etc.) here as in the previous cases. As displayed in Fig.~\ref{fig:uc_cavity_CaseI}, better accuracy is then achieved, which can also be seen in Table \ref{table:relative_errorcavity_inverse_uniformgrid}.

Based on the results in Figs. \ref{fig:cavity_inverse}-\ref{fig:uc_cavity_CaseI}, we can further conclude that: (1) the predictive accuracy can be enhanced by increasing the number of training data as we compare the predictions in Figs. \ref{fig:cavity_inverse} and \ref{fig:uc_cavity_CaseI} with the results in Fig. \ref{fig:uc_cavity_lines}, (2) the predictions of PINN-BGK are more accurate than those from regression, which again demonstrates the physics-informed constraint is able to improve the predictive accuracy, and (3) we can obtain better accuracy as we add more measurements near the vortex zone as well as the boundaries when comparing the results in Fig. \ref{fig:uc_cavity_CaseI} with those in Fig. \ref{fig:uc_cavity_lines}.

\section{Summary}
\label{sec:summary}

The Boltzmann equation with the Bhatnagar-Gross-Krook collision model (Boltzmann-BGK equation) is a well known model for describing multiscale flows from the hydrodynamic limit to free molecular flows. The recently proposed physics-informed neural networks (PINNs) have shown expressive power for solving both forward and inverse PDE problems in fluid mechanics and beyond. By leveraging the merits of the Boltzmann-BGK equation and PINNs, we have solved forward and inverse problems in multiscale flows via Boltzmann-BGK formulation using PINNs (PINN-BGK).
In particular, the proposed PINN-BGK contains three sub-networks, i.e., one for the equilibrium distribution function, a second one to approximate the non-equilibrium distribution function, and a third one to encode the Boltzmann-BGK equation as well as the corresponding boundary/initial conditions. By minimizing the residuals of all governing equations and the mismatch between the predicted and provided boundary/initial conditions, we can then solve both forward and inverse problems in multiscale flows with the same formulation and the same code.

We first test the PINN-BGK for solving forward problems in multiscale flows. Specifically, the PINN-BGK is applied to model both continuous  (e.g., Taylor-Green flows, cavity flows, etc.) and rarefied flows with Knudsen number ranging from 0.01 to 5. The results demonstrate that the PINN-BGK is capable of approximating accurately the solution to the Boltzmann-BGK equation given boundary/initial conditions. We then apply the PINN-BGK to solve inverse problems, i.e., inferring the velocity field based on a limited number of scattered interior observations on the velocity without any information on the boundary conditions. In particular, we test the performance of the PINN-BGK for micro Couette and cavity flows with a Knudsen number ranging from 0.01 to 10. Our results demonstrate that the PINN-BGK is capable of reconstructing the velocity field with good accuracy based on a small set of interior measurements on velocities. This is quite promising for multiscale flows within complicated geometries in which accurate boundary conditions are difficult to obtain, such as shale gas flow in pores at scales varying from $nm$ to $mm$. Finally, we demonstrate that we can achieve three-fold speedup by using  transfer learning as compared to the standard training, i.e., Adam optimizer plus L-BFGS-B, for the specific Kovasznay flow case we studied; we expect even greater speedup for three-dimensional time-dependent flows.


\section*{Acknowledgement}
Q. Lou would like to acknowledge the support of the National Natural Science Foundation of China (Grant No.51976128) and the Natural Science Foundation of Shanghai (Grant No. 19ZR1435700). X. Meng and G. E. Karniadakis would like to acknowledge the support of PhILMS grant DE-SC0019453.  The authors would like to thank Mr. C. Zhang for providing the DUGKS results for inverse problems.

\bibliographystyle{elsarticle-num}
\bibliography{refs}

\end{document}